\documentclass[12pt,preprint]{aastex}

\begin{document}

\def \lndobj {$\log N_{\rm D I} = 16.19 \pm 0.04$}
\def \lnhobj {$\log N_{\rm H I} = 20.67 \pm 0.05$}
\def \lndhobj {$\log \rm{D/H} = -4.48 \pm 0.06$}
\def \lndhallfull {$\log \rm{D/H} = -4.54581 \pm 0.03606$}
\def \lndhall {$\log \rm{D/H} = -4.55 \pm 0.04$}
\def \dhobh {$\Omega_{b}h^{2} = 0.0213 \pm 0.0013 \pm 0.0004$}
\def \etal {\textit{et al.}}
\def \zabs {$z_{\rm abs}$}
\def \zem {$z_{\rm em}$}
\def \lya  {Ly$\alpha$}
\def \Lyb  {Ly$\beta$}
\def \lyb  {Ly$\beta$}
\def \Lyg  {Ly$\gamma$}
\def \lyg  {Ly$\gamma$}
\def \ly5  {Ly-5}
\def \ly6  {Ly-6}
\def \ly7  {Ly-7}
\def \lyaf  {\lya\ forest}
\def \kms {km s$^{-1}$}
\def \nhi {$N_{\rm H I}$}
\def \ndi {$N_{\rm D I}$}
\def \lnhi {$\log N_{\rm H I}$}
\def \lndi {$\log N_{\rm D I}$}
\def \bhi {$b_{\rm H I}$}
\def \cmm {cm$^{-2}$}
\def \cmmm {cm$^{-3}$}
\def \da {$D_{A}$}
\def \btemp {$b_{temp}$}
\def \bturb {$b_{turb}$}
\def \ly {Lyman}
\def \mf {$\times 10^{-5}$}

\def \qfirst {PKS~1937--1009}
\def \qsecond {Q1009+2956}
\def \qthird {Q0130--4021}
\def \qfourth {HS~0105+1619}
\def \qfifth {Q1243+4037}
\def \qhst {PG~1718+4807}
\def \object{SDSS1558-0031}

\def \ob {$\Omega_{b}$}
\def \obh {$\Omega_{b}h^{2}$}
\def \ol {$\Omega_{\Lambda}$}
\def \olh {$\Omega_{\Lambda}h^{2}$}
\def \om {$\Omega_{m}$}
\def \omh {$\Omega_{m}h^{2}$}
\def \ETA {$\eta $}
\def \het {$^3$He}
\def \hef {$^4$He}
\def \lisv {$^7$Li}
\def \yp {Y$_p$}

\title{The Deuterium to Hydrogen Abundance Ratio Towards the QSO SDSS1558-0031\altaffilmark{1}}

\author{
JOHN M.\ O'MEARA\altaffilmark{2}, SCOTT BURLES\altaffilmark{2}, 
JASON X. PROCHASKA\altaffilmark{3,4}, GABE E. PROCHTER\altaffilmark{3,4}, REBECCA A. BERNSTEIN\altaffilmark{5}, and KRISTIN M. BURGESS\altaffilmark{6}
}
\altaffiltext{1}{This paper includes data gathered with the 6.5 meter Magellan Telescopes located at Las Campanas Observatory, Chile.}

\altaffiltext{2}{MIT Kavli Institute for Astrophysics and Space Research, 
Massachusetts Institute of Technology, Cambridge, MA 02139, 
{\tt omeara@mit.edu, burles@mit.edu}}

\altaffiltext{3}{Visiting Astronomer, W.M. Keck Observatory which
is a joint facility of the University of California, the California
Institute of Technology and NASA}

\altaffiltext{4}{Department of Astronomy and Astrophysics,
UCO/Lick Observatory;
University of California, 1156 High Street, Santa Cruz, CA 95064, {\tt xavier@ucolick.org, prochter@ucolick.org}}

\altaffiltext{5}{Department of Astronomy, University of Michigan, Ann Arbor, MI 48109}

\altaffiltext{6}{Department of Physics, Princeton Univeristy, Princeton, NJ 08544, {\tt kburgess@astro.princeton.edu}}

\begin{abstract}

We present a measurement of the D/H abundance ratio in 
a metal-poor damped \lya\ (DLA) system along
the sightline of QSO \object.  The DLA system is at redshift 
$z = 2.70262$, has a neutral column density of \lnhobj\ \cmm,
and a gas-phase metallicity [O/H]$= -1.49$ 
which indicates that deuterium astration is negligible.
Deuterium absorption is 
observed in multiple Lyman series with a column density of 
\lndobj\ \cmm, best constrained
by the deuterium Lyman-11 line.  
We measure \lndhobj, which when combined with 
previous measurements along QSO sightlines gives a best estimate of 
\lndhall\, where the 1$\sigma$ error estimate comes from a
jackknife analysis of the weighted means.  
Using the framework of standard big bang nucleosynthesis, this 
value of D/H translates into a baryon density of \dhobh, where the error terms represent the $1\sigma$ errors from D/H and the uncertainties in the nuclear reaction rates respectively.  
Combining our new measurement with previous measurements 
of D/H, we no longer find compelling evidence for a trend of D/H with \nhi.
\end{abstract}


\section{INTRODUCTION}
For the last decade, measurements of the primordial D/H ratio in QSO sightlines have provided increasingly more precise constraints on the cosmological baryon density.  Although the measurement of D/H is simple in principle, 
compared to the other light elements produced during big bang nucleosynthesis, finding those QSO absorption lines systems which are suitable for measuring D/H has proven observationally challenging 
\citep{tytler_review}.

For a QSO absorption system to show D/H, a number of criteria must be met (see \citet{kirkman03}, hereafter K03,  for a more detailed discussion).  First, the hydrogen column density must be large enough (since D/H is of the order of one part in $10^5$) such that deuterium can be observed using modern high-resolution spectrographs.  Second, the velocity structure of the hydrogen absorption must be simple enough, ideally a single component of gas, so that the deuterium absorption is well resolved given the small 82 \kms\ offset 
from the hydrogen Lyman lines.  
Third, there can be little to no interloping \lya\ forest or 
metal lines at the position of the deuterium absorption, since such absorption strongly complicates attempts to constrain the deuterium column density.  Unfortunately, \lya\ forest absorption is both ubiquitous and stochastic in high redshift QSO spectra.  Finally, the background QSO must be bright enough to obtain high signal-to-noise, high-resolution spectroscopy at $\lambda < 4000$\AA\
with a reasonable allotment of telescope time.  Each one of these criteria act to decrease the probability that a D/H measurement can be made towards any given QSO, and since \textit{all} the criteria must be met, the resultant probability of a QSO sightline being suitable for measuring D/H is very low, with only approximately 1\% of QSOs at $z \simeq 3$ able to provide a measurement of D/H.

To date, there are few measurements of 
D/H in QSO spectra (\citet{bt98a},\citet{bt98b}, \citet{omeara01}, \citet{pettinibowen}, K03, \citet{levshakovDH}, \citet{crighton04}).  These measurements constrain the baryon density \obh\ through the framework of standard big bang nucleosynthesis (SBBN), which predicts the abundances of the light elements as a function of the baryon--to--photon ratio $\eta$ and the expansion rate of the universe \citep{kolb}, 
and through the cosmic microwave background radiation, which provides the photon density.  A measurement of the ratio of any of the light nuclei produced in SBBN gives the baryon density, and  measurement of additional abundance ratios test the theory (see \citet{steigman05},  \citet{pettinidh}, and references therein for a current census of D/H and the other light element abundances).

Recent measurements of the temperature angular power spectrum of the CMB (Spergel \etal\ 2006) also provide a measurement of \obh\,
depending on the assumptions made, 
with a level of accuracy roughly equal to or greater than that provided by D/H.
Nevertheless, measurements of D/H are still important for a number of reasons. 
First, primordial D/H probes the universe at one of the earliest times in the universe accessible with current observational and theoretical techniques.  
Second, the light element abundances predicted from SBBN do not all agree with each other; most notably the observationally inferred abundance of \lisv\ is significantly lower than that expected from SBBN and D/H \citep{fields06}. 
Third, D/H can help constrain deviations from SBBN, such as inhomogeneous 
BBN \citep[e.g.][]{lara06}, 
relic primordial particle decays \citep[e.g.][]{jedamzik04},
 or non-standard neutrino physics \citep[e.g.]{abazajian05}.  
Fourth, the dispersion in the measurements of D/H is larger than would be expected from the individual measurement errors (K03), i.e., the data demand both a better understanding of the errors on the current measurements and new constraints. 
Fifth, the value of \ob\ derived from D/H and SBBN requires many fewer priors than the CMB derived value.  Moreover, the \ob\ from D/H can be used in principle as a prior in the CMB analysis, and the ratio of the values for \ob\ from D/H and the CMB offer a precision test of the hot Big Bang model.
Finally, the existing measurements of D/H show evidence of a trend of 
decreasing D/H with increasing \nhi.  
K03 suggested that this trend 
(and the dispersion in D/H values) is due to error under-estimation.

Fortunately, the Sloan Digital Sky Survey \citep[SDSS,][]{sdssdr4}, 
by virtue of its large sample of high redshift QSO spectra 
(the Data Release 4 alone contains 5036 QSOs with $z > 2.7$) 
gives us a new data set to find those special sightlines which can show D/H. 
In this \textit{Letter}, we present a new measurement of D/H in a 
QSO sightline from the SDSS, \object, which was chosen as part of 
our high-resolution survey for Lyman limit absorption 
\citep{omeara06}.

\section{OBSERVATIONS}
We have obtained two high-resolution spectra of the $z=2.83$ quasar \object\ using two different spectrographs, the MIKE \citep{bernstein03} 
echelle spectrograph on the 6.5 meter Magellan Clay telescope at Las Campanas, 
and the upgraded HIRES \citep{vogt94}
spectrometer on the 10 meter Keck-I telescope on Mauna Kea.  
The MIKE spectrum was obtained as part of our ongoing high-resolution 
survey for Lyman Limit absorption and was selected from the SDSS 
because of the redshift and brightness of the QSO.  The MIKE spectrum 
was obtained on May 10, 2004 and covers the spectral range 
3221--7420 \AA, with an exposure time of 3600 seconds.  The data were obtained in sub-arcsecond seeing with a one arcsecond slit, which provides a resolution of $R=28,000$ and $R=22,000$ on the blue and red arms of the spectrograph respectively.
The MIKE data was reduced using the MIKE reduction pipeline\footnote{http://web.mit.edu/$\sim$burles/www/MIKE/}, and has a signal to noise ratio of approximately 12 at $\lambda = 4000$\AA.

The HIRES spectrum was taken on 2006 April 11 
and covers the spectral range 3338--6200 \AA, with an exposure time of 4100 seconds.  The data were taken in sub-arcsecond seeing with a 1.148 arcsecond slit, which provides a resolution of $R=34,000$. The data were reduced using the HIRES reduction pipeline\footnote{http://www.ucolick.org/$\sim$xavier/HIRedux/index.html},
and have a signal to noise ratio of approximately 20 at $\lambda = 4000$ \AA.
The HIRES data are the primary source for the measurement of D/H presented below owing to the higher signal-to-noise and spectral resolution.  
With the exception of the H~I \lya\ line, we used the HIRES spectrum to determine all values for column densities presented in the text.  Because 
we were more successful at fluxing data from the MIKE spectrometer, 
we use the flux calibrated MIKE spectrum to constrain the \nhi\ value
in the \lya\ line whose profile spans several echelle orders.

\section{Analysis}

Inspection of the MIKE spectrum of \object\ shows that there is a DLA at $z = 2.70262$ which is also responsible for the break in flux from the Lyman limit of the absorber at $\lambda \approx 3885$ \AA.  
The parameters which describe the observed, single component of 
absorption ($N$,$b$,$z$) for the Lyman series and metal-lines
in the DLA are presented in Table~\ref{dhlinetab}.  
The parameters and their errors are derived
predominately from Voigt profile fits to the data using the VPFIT 
routine kindly provided by R. Carswell and J. Webb.  
For some of the metal-line 
transitions, we opt instead to use the apparent 
optical depth technique \citep{savage91} to measure the column densities, 
and then use that column density as a fixed input parameter to 
VPFIT to determine the $z$ and $b$ values for the absorption in question.  
The absence of metal-line absorption at the position of D
aruges that the observed feature in the Lyman series is not interploping H.

\begin{figure}
\begin{center}
\includegraphics[scale=0.7]{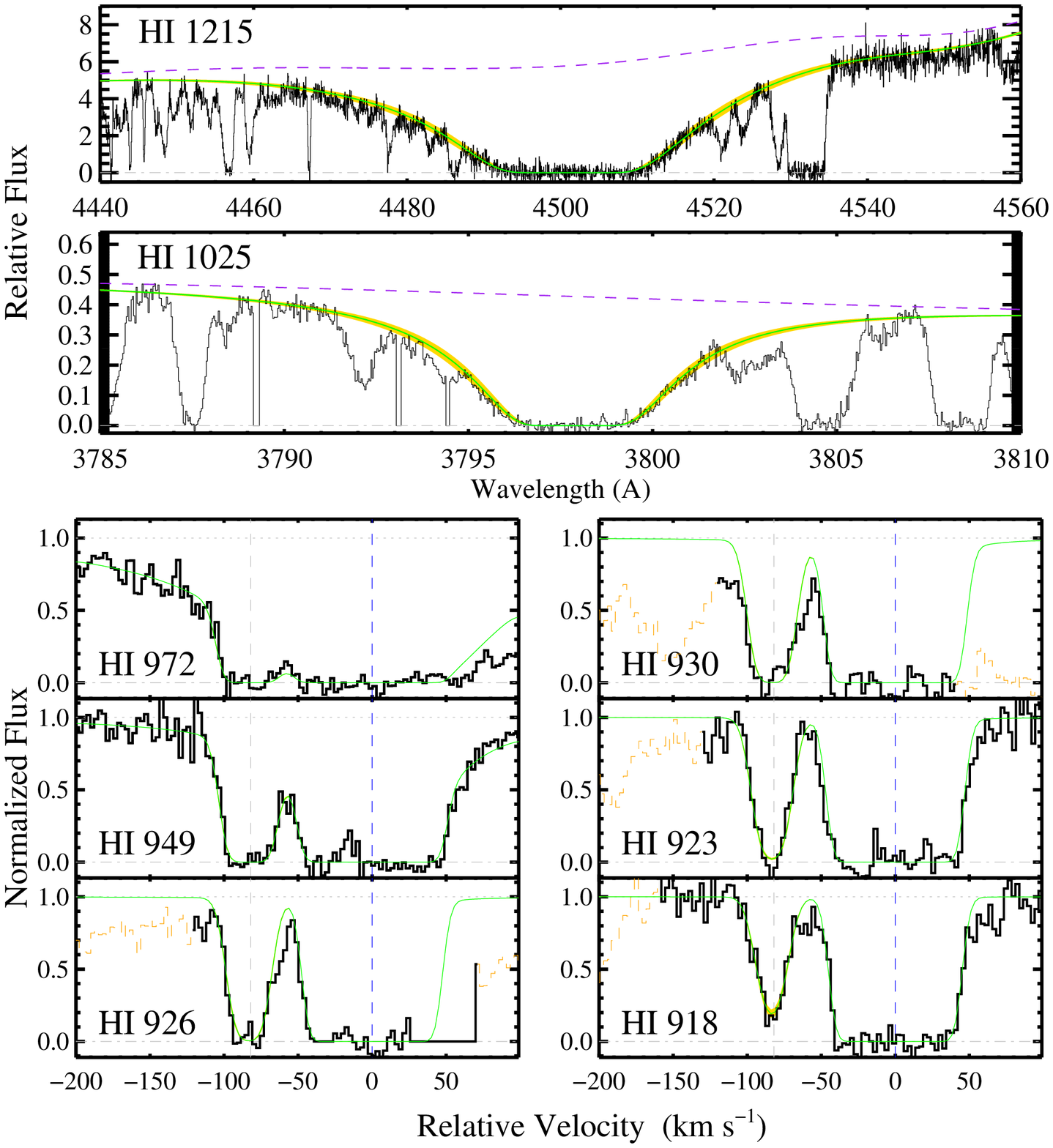}
\caption{H~I and D~I Lyman series absorption in the $z = 2.70262$ DLA towards \object.  The \lya\ transition (top panel) comes from our MIKE spectrum of \object\ and the remaining panels come from our HIRES data.  For the \lya\ and \lyb\ transitions, we present unnormalized data, along with a dashed line which shows our estimate for the local continuum level.  The remaining Lyman series transitions are shown continuum normalized.  The solid green line shows the best single-component fit to the D~I and H~I absorption. We derive our estimates of the \nhi\ from the damping wings present in Lyman $\alpha$--$\delta$, and the \ndi\ from the unsaturated D~I Lyman-11 transition ($\sim $918 \AA\ rest wavelength). }
\label{fig_lyseries}
\end{center}
\end{figure}

The hydrogen column density of the DLA is sufficiently large
to show damping wings in the Lyman $\alpha$--$\gamma$ transitions 
(Figure~\ref{fig_lyseries}).  
These features allow for a precise measurement of the H~I column density.  
In the case of \object\, the DLA \lya\ lies near the QSO emission line and the
assignment of the continuum level over the full 
extent of the absorption feature is 
non-trivial.  
Fortunately, this issue is minimized in two ways.  First, the MIKE spectrum is flux calibrated, which allows for an easier assignment of the continuum level to the \lya\ line, although it is still subject to unidentified emission features
inherent to the QSO.  
Second, the H~I derived from the damping features in the higher order lines is less susceptible to large continuum shape errors, because the profile spans a 
significantly smaller wavelength range than the \lya\ line.  In particular, 
the \lyb\ line of the DLA places an excellent constraint on the \ion{H}{1}
column density, 
because the line has prominent damping features, covers only $\approx 35$ \AA, and has little interloping hydrogen absorption.

To arrive at the best estimate of \nhi\ we simultaneously vary the values of \nhi\ along with the shape and amplitude of the local continuum level.  This variation continues until we arrive at a value of the \nhi\ which best reproduces the data whilst having a reasonable continuum shape.  
We adopt a redshift and velocity width of the H~I, $z=2.702646 \pm 0.000010$ and $b=13.56 \pm 1.0$, from a fit to the the higher order Lyman series transitions.
Of some concern is the fact that the redshift of the H~I agrees only at the $\approx 3$ \kms\ level with the redshift inferred from low-ion metal-lines (Table~\ref{dhlinetab}).  We note, however, that there exists some degeneracy between  $b$  the $z$ for the Lyman series transitions we use, along with the increasing effects of poor signal-to-noise for shorter wavelength data 
(i.e. higher up the Lyman series).  Furthermore, we cannot discount the possibility that the H~I gas is multi-component, however there is little evidence 
from the metal-line transitions that this is the case. 
Furthermore the Lyman series lines
all appear to be well fit using a single component,
with a few departures due to interloping 
\ion{H}{1} gas at different redshifts from the system which shows deuterium. 
When we consider the Lyman $\alpha$--$\gamma$ transitions, we arrive at a best estimate of the H~I of \lnhobj\ \cmm. 
The errors on \nhi\ are dominated by continuum uncertainties and by signal-to-noise, two effects which correlate, particularly on smaller wavelength scales.

As can be seen in Figure \ref{fig_lyseries}, we observe resolved
absorption by deuterium in the 
Lyman series from \lyg\ all the way through to Lyman-13.  
Because the D~I column density is large, 
the absorption is saturated until we reach deuterium Lyman-11, 
which offers the best constraint on the \ndi\ value.    
For this transition, we measure \lndobj\ \cmm, where the errors come from the error estimate of VPFIT alone (i.e. independent of continuum error).  The transition suffers from mild contamination by 
interloping hydrogen on $\approx 25$ \kms\ to the red side of the absorption profile, but this absorption has little effect on the \ndi\ value.
Fits to the data including and excluding a model for the interloping 
hydrogen improve the $\chi^2$ for the fit without changing the value
or uncertainty in the \ndi\ value.
We have also estimated \ndi\ using the AODM technique, and 
arrive at a consistent value \lndi\ $=16.20 \pm 0.04$\cmm.
The optical depth of this absorption feature is ideal for measuring a 
column density because it is highly insensitive to the local continuum 
level placement.  If we vary the amplitude of the continuum level by 
as much as 20\% about the adopted value the central value, of 
\ndi\ changes by less than the statistical error.  
The \ndi\ value is further constrained by 
other Lyman series lines, e.g.\  the depth of the deuterium 
Lyman--8 transition rules out significantly larger or smaller values of \ndi.

We obtain a value of $z=2.702626 \pm 0.000007$ 
for the deuterium absorption, consistent with 
that of the H~I and other metals. 
We measure a velocity width of $b=10.48 \pm 0.78$ \kms\ for the deuterium absorption.  
Neutral hydrogen 
gas with \lnhi\ $\simeq 16.2$ \cmm\ is not expected to have such a narrow
Doppler parameter \citep{kt97} whereas the value is reasonable for D.

\begin{figure}
\begin{center}
\includegraphics[scale=0.65,angle=90]{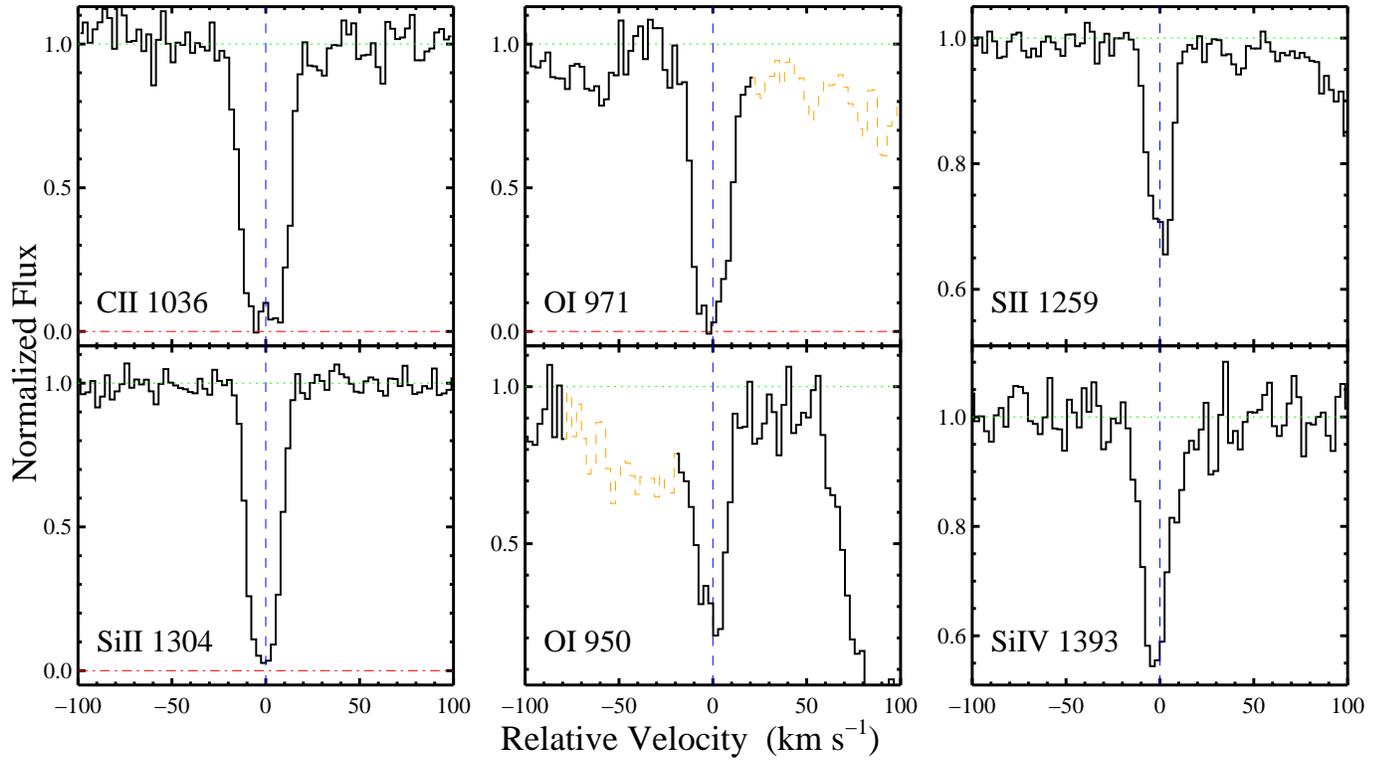}
\caption{Selected metal lines in the DLA towards \object.  The zero-point velocity corresponds to a redshift of $z=2.70262$.}
\label{fig_velp}
\end{center}
\end{figure}

We detect over 30 metal-line transitions in the DLA absorber.  
A subset of these are summarized in 
Table~\ref{dhlinetab} and are shown in Figure~\ref{fig_velp}.
In particular, we note the presence of \ion{O}{1} absorption 
at $z=2.702610 \pm 0.000005$,   
which is well described by a single component.  
\ion{O}{1} absorption is important for measuring D/H in that \ion{O}{1} 
directly traces the \ion{H}{1} gas \citep[see][]{omeara01}, 
and because \ion{O}{1}/\ion{H}{1} $\approx$ O/H in 
most environments (especially DLA).  
Adopting the measured value of $\log N_{OI} = 15.86$, we establish a 
metallicity of [O/H]$= -1.49$ for the absorber assuming 
the solar (atmospheric) oxygen abundance reported by \cite{asplund04}.
This metallicity is higher than all the other extragalactic measurements 
of D/H.  Nevertheless,  a 3\% solar metallicity 
implies minimal astration of D and we believe this system 
is still representative of primordial gas (see Figure 20 of K03 and \cite{romano06}).  
Finally, we note that the velocity structure of the absorber, 
as traced by the metal lines, is amongst the simplest yet 
observed for a DLA \citep{pw01}.

\section{Discussion}
We now discuss the value of D/H we obtain for the absorber and place it within the context of the combined D/H ratio for all QSO absorption systems.
The best estimates of the \ion{H}{1} and \ion{D}{1} column densities 
in the DLA towards \object\
imply a value of \lndhobj. The errors on D/H stem primarily from the effect of continuum placement uncertainty on the H~I column density, and the signal-to-noise ratio of the data at Lyman-11 where the \ion{D}{1} column density is best constrained.

\begin{figure}
\begin{center}
\includegraphics[scale=0.7]{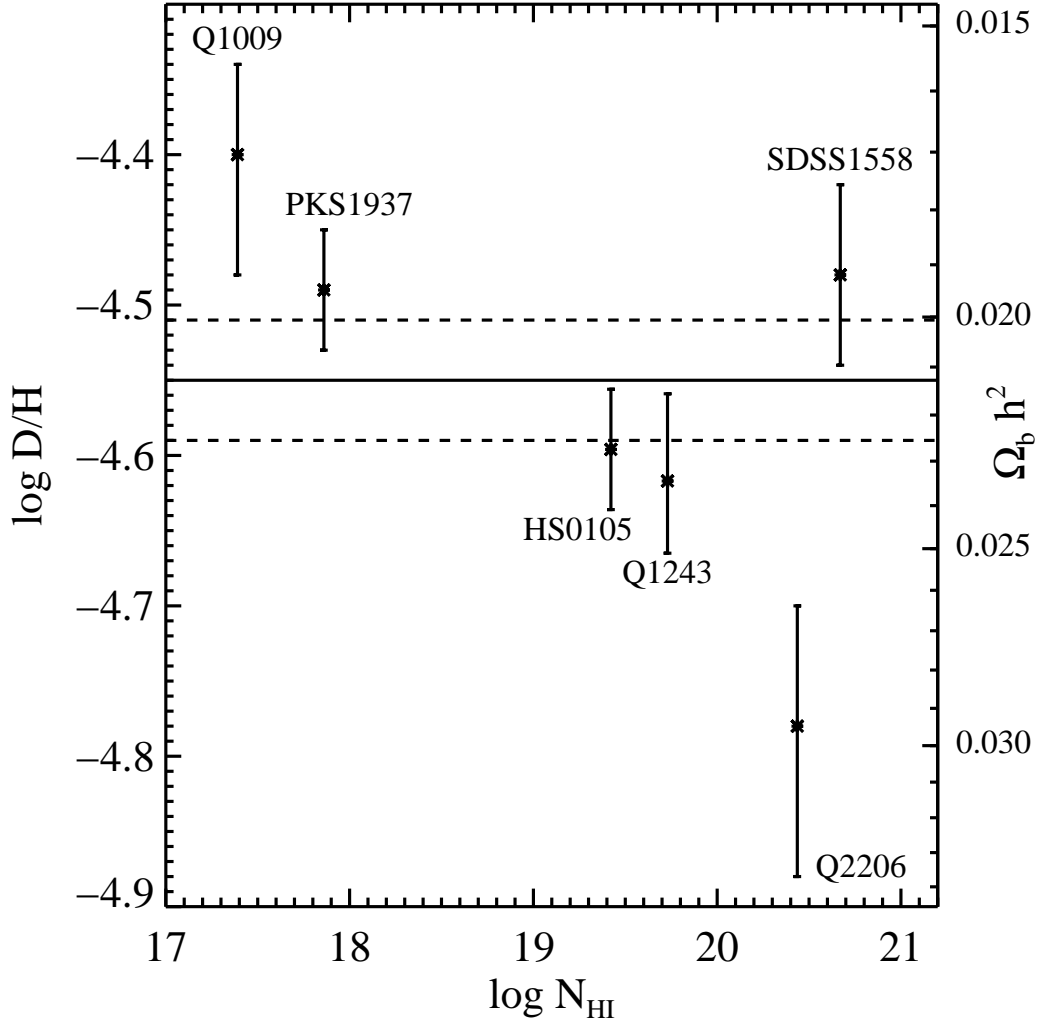}
\caption{Values of the D/H ratio vs. \lnhi.  With the new value for D/H from \object, a linear trend of D/H with \nhi\ is no longer statistically significant nor a good description of the data.  The  horizontal lines represent the weighted mean and jackknife errors of the 6 measurements, \lndhall. The right hand axis shows how the values of D/H translate into values for \obh\ using SBBN. 
}
\label{fig_alldh}
\end{center}
\end{figure}

Turning now to the combined D/H value from QSO sightlines, Figure \ref{fig_alldh} shows the new value of D/H from \object\ along with the previous values of D/H taken from the sample discussed in K03.  We do not include the result of Crighton \etal\ (2004), since we feel that the errors on D/H in this system have been under-estimated, particularly 
for the reported \nhi\ value. 
We do not include the results of \citet{levshakovDH} for the reasons given 
in K03.
The horizontal solid line shows the value for the weighted mean 
of the data  
\lndhallfull, which we round to 
$\log \rm{D/H} = -4.55 \pm 0.04$
to keep consistent with the literature, 
and the dashed lines show the $\pm 1 \sigma$ uncertainties 
estimated from a jackknife analysis of the weighted means. In the case of asymmetric errors on individual D/H measurements, we have
adopted the larger of the errors for the calculation of the weighting. 

Prior to the addition of \object\ to the sample of D/H measurements of K03, a $\chi^2$--minimizing linear fit to the data of the form 
$\log \rm{D/H} = -2.914 \pm 0.467 - (0.087 \pm 0.025)\times$\lnhi\ provided an acceptable fit to the data ($P_{(\chi^2 > \chi^2_{\rm{fit}})} = 0.74$).  With the inclusion of \object, however, we see a significant decrease in the 
likelihood that there is a D/H trend with \nhi.  
Although the data are best fit with a non--zero slope,
$\log \rm{D/H} = -3.707 \pm 0.385 - (0.044 \pm 0.021)\times$\lnhi,
the slope differs from zero at only the $2\sigma$ level (even before
accounting for the presence of \nhi\ in both axes).
Furthermore, this is not a good model of the data, with $P_{(\chi^2 > \chi^2_{\rm{fit}})} = 0.04$.  As such, the data give little confidence to the existence of a trend of D/H with \nhi.  
Finally, if we were to include 
the \cite{crighton04} and \cite{levshakovDH} results, the 
likelihood that D/H depends on \nhi\ is further diminished.

The 
presumption of a single value for D/H, however, is still not supported by the observations; the observed scatter exceeds that expected assuming 
the error estimates reported in the literature.  Our adopted error on 
the best estimate for D/H from the jackknife estimation exceeds the 
error on the weighted mean by a factor of two.
Likewise, the $\chi^2$ of the six measurements of D/H about the weighted 
mean value of \lndhall\ is high, 
with $P_{(\chi^2 > \chi^2_{\rm{D/H}})} = 0.01$.  Although there is the possibility that the scatter in the individual D/H measurements is real, we prefer
the hypothesis of K03 that the errors in some of the individual measurements, if not all of them, are underestimated.  It is likely that a combination of new methods of error analysis and new QSO  sightlines is required to fully address the excess scatter in the D/H measurements.

Using the framework from SBBN \citep{bnt2001}, the value for D/H of \lndhall\ translates to a value for the cosmological baryon density of \dhobh, where the first error term comes from the errors on D/H explained above, and the second term from the uncertainties in the nuclear reaction rates.  By comparison, the WMAP three year result provides an estimate of \obh $=0.0223^{+0.0007}_{-0.0009}$, which lies  within the $1\sigma$ error estimate on \obh\ from D/H (Spergel \etal\ 2006).

Finally, we note that the absorber showing D/H in the spectrum of \object\ was discovered serendipitously as part of our survey for Lyman limit absorption, and is the first D/H measurement from a QSO first discovered by the SDSS.  
The SDSS Data Release 3 alone has 405 DLA with 
redshifts optimal for D/H \citep[$2.51 \le z \le 4.0$][]{dla_dr3}, 
Assuming a small fraction of these DLA provide measurements of D/H, 
the SDSS will give many tens of measurements.  The situation improves further
if one considers the Lyman limit systems toward the SDSS QSO sample.
This contrasts with the SLLS and DLA which give the \nhi\ from the \lya\ and \lyb\ lines, and the \ndi\ from the unsaturated D~I lines in the Lyman series.  Because of this effect, the DLA and SLLS offer more pathlength per QSO to potentially find D/H.  All of these effects combine to give a likely distribution of D/H measurements which is roughly independent of \nhi, a hypothesis which is already being hinted at in the current sample, since two measurements come from LLS, two from SLLS, and two from DLA.  
Altogether, the SDSS offers the best opportunity for investigating 
the larger than expected scatter in D/H and correlations with
\nhi, metallicity, etc. 

\acknowledgments

The authors wish to recognize and acknowledge the very significant
cultural role and reverence that the summit of Mauna Kea has always
had within the indigenous Hawaiian community.  We are most fortunate
to have the opportunity to conduct observations from this mountain.
We thank M. Pettini, G. Steigman, and the referee, P. Molaro, for inciteful questions and comments which improved this letter.
GEP and JXP are supported by NSF grant AST-0307408.  
JO and SB acknowledge support from NSF grant AST-0307705.


\begin{deluxetable}{llcccc} \tablecaption{\label{dhlinetab} IONS OBSERVED IN THE $z = 2.70262$ DLA TOWARDS \object} \tablehead{ \colhead{Ion} & \colhead{log N\tablenotemark{a}} & \colhead{$b$} & \colhead{$z$} &  \colhead{[X/H]\tablenotemark{b}} &
\colhead{Transition $\lambda_{\rm{r}}$}\\ 
\colhead{}    & \colhead{\cmm}  & \colhead{\kms} & \colhead{} & \colhead{} &\colhead{\AA} } 
\startdata 
H~I & $20.67\tablenotemark{c} \pm 0.05$ & $13.56 \pm 1.00$ & $2.702646 \pm 0.000010$ & --- & 1215.67, 1025.72, 972.54, 949.74 \\
D~I & $16.19\tablenotemark{c} \pm 0.04$ & $10.48 \pm 0.78$ & $2.702626 \pm 0.000007$ & --- & 917.88 \\ 
C~II & $>14.43\tablenotemark{d}$ & $9.69 \pm 0.21$ & $2.702611 \pm 0.000003$  & $>-2.63$ & 1334.53, 1036.34 \\ 
C~III & $>13.8 $ & --- & $2.702671 \pm  0.000054$ & --- & 977.02 \\ 
C~IV & $13.05 \pm 0.04$ & $9.00 \pm 1.35$ & $2.702614 \pm 0.000010$ & --- & 1548.20 \\ 
N~I & $14.46 \pm 0.02$ & $9.20 \pm 0.60$ & $2.702601 \pm 0.000005$ & -2.18 & 1134.41, 1134.17 \\ 
N~II & $13.71 \pm 0.04$ & $9.56 \pm 1.13$ & $2.702660 \pm 0.000008$ & --- & 1083.99 \\ 
O~I & $15.86 \pm 0.10$ & $6.38 \pm 0.40$ & $2.702610 \pm 0.000005$ & -1.49 & 971.74, 950.88 \\ 
Al~II & $12.73 \pm 0.05$ & $6.87 \pm 0.55$ & $2.702613 \pm 0.000005$ & -2.43 & 1670.79 \\ 
Si~II & $14.24\tablenotemark{d} \pm 0.01$ & $7.18 \pm 0.14$ & $2.702604 \pm 0.000002$ & -1.99 &1526.71,  1304.37 \\ 
Si~III & $>13.28 $ &  ---  & $2.702672 \pm 0.000003 $ & --- & 1206.50 \\ 
Si~IV & $12.72 \pm 0.04$ & $8.20 \pm 0.97$ & $2.702595 \pm 0.000008$ & --- & 1393.76 \\ 
S~II & $14.07 \pm 0.02$ & $7.05 \pm 0.47$ & $2.702626 \pm 0.000003$ & -1.74 & 1259.52 \\ 
Fe~II & $14.11 \pm 0.03$ & $6.47 \pm 0.32$ & $2.702605 \pm 0.000002$ &  -2.06 & 1608.45, 1144.94 \\ 
\enddata \tablenotetext{a}{Unless stated otherwise,  the errors on the column density are from VPFIT alone.}
\tablenotetext{b}{[X/H] $\equiv$ log(X/HI) - log(X/H)$_\odot$ where we have considered
only low-ion transitions and have not adopted ionization or depletion corrections.}
\tablenotetext{c}{See text for detailed description of the values and errors for this ion.} \tablenotetext{d}{Column density measurement from the apparent optical depth method.} \end{deluxetable}

\end{document}